\documentclass[aps,prd,11pt,twoside,tightenlines,nofootinbib,showpacs,preprint,superscriptaddress]{revtex4-1}
\usepackage[colorlinks=true]{hyperref}
\usepackage{xcolor}
\hypersetup{colorlinks=true, citecolor=blue, urlcolor=blue, linkcolor=blue}
\usepackage{amsmath,amssymb}
\usepackage[final]{graphicx}
\usepackage{subcaption} 
\usepackage{float}

\begin{document}
\title{Investigation of the $D^{0} \rightarrow K_S^{0} \pi^{0} \eta,\ K_S^{0} \pi^{0} \pi^0$ decays}

\author{Wei Liang}
\email{212201014@csu.edu.cn}
\affiliation{School of Physics, Hunan Key Laboratory of Nanophotonics and Devices, Central South University, Changsha 410083, China}

\author{Chu-Wen Xiao}
\email{xiaochw@gxnu.edu.cn (corresponding author)}
\affiliation{Department of Physics, Guangxi Normal University, Guilin 541004, China}
\affiliation{Guangxi Key Laboratory of Nuclear Physics and Technology, Guangxi Normal University, Guilin 541004, China}
\affiliation{School of Physics, Hunan Key Laboratory of Nanophotonics and Devices, Central South University, Changsha 410083, China}

\author{Guo-Mei Gan}
\affiliation{School of Physics and Telecommunication Engineering, Yulin Normal University, Yulin 537000, China}

\author{Shi-Qi Zhou}
\affiliation{School of Physics, Hunan Key Laboratory of Nanophotonics and Devices, Central South University, Changsha 410083, China}

\begin{abstract}

Inspired by the invariant mass distributions for the decays $D^{0} \rightarrow K_S^{0} \pi^{0} \eta$ and $D^{0} \rightarrow K_S^{0} \pi^{0} \pi^{0}$ reported by the BESIII Collaboration, we investigate these processes with an unified final state interaction formalism by incorporating both the $S$-wave pseudoscalar meson-pseudoscalar meson interactions within a chiral unitary approach and the $P$-wave contribution from the intermediate resonance $\bar{K}^*(892)$. 
By performing a combined fit to the invariant mass spectra and taking into account the coherence between the $S$- and $P$-waves, our results are in agreement with the experimental data. 
For the decay $D^{0} \rightarrow K_S^{0} \pi^{0} \eta$, the structure near 1.0 GeV in the $\pi^0 \eta$ invariant mass distribution corresponds to the signal of the $a_{0}(980)$, which is dynamically generated from the $S$-wave interactions. 
In the case of $D^{0} \rightarrow K_S^{0} \pi^{0} \pi^{0}$, the near-threshold enhancement in the $\pi^0\pi^0$ mass distribution arises from the combined contributions of the $f_0(500)$ and the intermediate $\bar{K}^*(892)$, while the cusp-like structure around 1 GeV$^2$ is associated with the $f_0(980)$.

\end{abstract}

\maketitle


\section{Introduction}

The study of strong-interaction dynamics in the low-energy region of particle physics remains a core challenge in understanding the nonperturbative nature of quantum chromodynamics (QCD)~\cite{Cheng:2010cb,Cheng:2010ry,ParticleDataGroup:2024cfk,Chen:2016qju,Hosaka:2016pey,Guo:2017jvc}. 
In this regime, the QCD coupling constant becomes large, rendering perturbative methods unreliable. 
To reveal the non-perturbative QCD behaviour, the hadronic decays of charm mesons provide a valuable opportunity to investigate the interplay between the short-distance weak interactions and the long-distance strong dynamics~\cite{Esposito:2016noz,Olsen:2017bmm,Yuan:2018inv,Brambilla:2019esw,Wang:2021ail,Li:2024rqb}. 
In particular, the final state interaction among two of the outgoing pseudoscalar mesons for three-body charm meson decays can generate abundant resonance structures in the invariant mass spectra, which not only provide insights into the weak decay mechanisms of charm quarks but also serve as sensitive probes of the complex dynamics underlying final state strong interactions.

Among the long-standing open issues in light-hadron spectroscopy, the nature of the scalar states $f_0(500)$, $f_0(980)$, and $a_0(980)$ remains one of the most intriguing and enduring puzzles, see more discussions in the review of the meson resonances below 1 GeV in the Particle Data Group (PDG)~\cite{ParticleDataGroup:2024cfk} and the discussions in the works~\cite{Close:2002zu,Debastiani:2016ayp,Liang:2016hmr,Wang:2021naf,Feng:2020jvp,Wang:2020pem}. 
Given their unusual properties, especially the proximity of the states $f_0(980)$ and $a_0(980)$ to the $K\bar{K}$ threshold, these states have been interpreted within various theoretical frameworks, including conventional $q\bar{q}$ mesons, compact tetraquarks, hadronic molecules, and hybrids involving multiple components~\cite{Godfrey:1985xj,Morgan:1993td,Tornqvist:1995ay,Lee:2022jjn,Jaffe:1976ig,Zou:1994ea,Janssen:1994wn,Oller:1997ti,Locher:1997gr,Baru:2003qq,Albuquerque:2023bex}. 
Against this backdrop, the chiral unitary approach offers a natural and successful theoretical framework, where the low-lying scalar resonances emerge dynamically from the unitarized $S$-wave interactions of coupled pseudoscalar-meson channels, rather than being introduced ad hoc as elementary Breit-Wigner resonances~\cite{Oller:1997ti,Oset:1997it,Toledo:2020zxj,Oller:2000fj,Guo:2005wp}.

Experimentally, the BESIII, BaBar, Belle, and CLEO Collaborations reported many experimental measurements about three-body decays of the charm hadrons, such as in the Refs.~\cite{BaBar:2010nhz,LHCb:2019tdw,BESIII:2020pxp,Belle:2021dfa,BESIII:2019xhl,CLEO:2008icw,BESIII:2024tpv}, which provide a good opportunity to investigate these light scalar mesons~\cite{Wang:2021kka,Liang:2014ama,Xie:2014gla,Ling:2021qzl,Ding:2024lqk,Ding:2023eps,Dai:2018rra} with their results on the two-body spectra of the final states $\pi \pi$, $\pi \eta$, and/or $K \bar{K}$. 
Recently, the BESIII Collaboration reported their precise measurements of the decays $D^0 \rightarrow K_S^0 \pi^0 \eta$~\cite{BESIII:2025wmd} and $D^0 \rightarrow K_S^0 \pi^0 \pi^0$~\cite{BESIII:2025sea}. 
In Ref.~\cite{BESIII:2025wmd}, they performed an amplitude analysis of the decay $D^0 \rightarrow K_S^0 \pi^0 \eta$ and obtained the branching fraction $\mathcal{B} (D^0 \rightarrow K_S^0 \pi^0 \eta)=(1.016 \pm 0.013_{stat} \pm 0.014_{syst})\%$, together with $\mathcal{B}(D^0 \rightarrow K_S^0 a_0(980)^0, a_0(980)^0 \rightarrow \pi^0 \eta =(9.88 \pm 0.37_{stat} \pm 0.42_{syst}) \times 10^{-3} $, showing that the contribution from the resonances $a_0(980)^0$ and $\bar{K}^*(892)$ plays a dominant role in the decay procedure. 
This decay channel was first observed by the CLEO Collaboration~\cite{CLEO:2004umu}, where the results with a Dalitz-plot analysis indicated the dominant contributions from the sub-decays $K_S^0 a_0(980)$ and $K^*(892) \eta$. 
In Ref.~\cite{BESIII:2025sea}, the BESIII Collaboration reported the measurement results of the decay $D^0 \rightarrow K_S^0 \pi^0 \pi^0$ with an amplitude analysis using 20.3 fb$^{-1}$ of $e^+e^-$ data collected at $\sqrt{s}=3.773$ GeV, where they measured the branching ratio as $\mathcal{B}(D^0 \rightarrow K_S^0 \pi^0 \pi^0)=(1.026 \pm 0.008_{stat} \pm 0.009_{syst})\%$, and found that the dominant intermediate process was $D^0 \rightarrow \bar{K}^*(892)^0 \pi^0$ with significant contribution from the $S$-wave component. 
For this decay channel, the CLEO Collaboration had also performed a Dalitz analysis and found that the data required contributions associated with the resonances $K^*(892)$, $f_0(980)$, and the broad one $\sigma/f_0(500)$~\cite{CLEO:2011cnt}. 

On the theoretical side, the decays $D^0 \rightarrow K_S^0 \pi^0 \eta$ and $D^0 \rightarrow K_S^0 \pi^0 \pi^0$ are particularly attractive to investigate the resonance properties of these scalar states $f_0(500)$, $f_0(980)$, and $a_0(980)$, since the $\pi^0 \eta$ system is an ideal channel to observe the isovector state $a_0(980)$, while the $\pi^0 \pi^0$ system filters isoscalar configurations and is sensitive to the $f_0(500)$ and $f_0(980)$. 
Indeed, using a final state interaction framework based on the chiral unitary approach, the decays $D^0\to K_S^0 f_0(500)$, $D^0\to K_S^0 f_0(980)$, and $D^0\to K_S^0 a_0(980)$ were studied in Ref.~\cite{Xie:2014tma} with a weak interaction diagram of dominant $W$-internal emission, where they predicted that the signals of these scalar resonances could be found in the spectra of the $\pi^+ \pi^-$ and $\pi^0 \eta$ in the final states $\bar{K}^0 \pi^+ \pi^-$ and $\bar{K}^0 \pi^0 \eta$, respectively. 
With an analogous formalism, the decay $D^0 \rightarrow \bar{K}^0 \pi^0 \pi^0$ was investigated in Ref.~\cite{Zhang:2024myn} with the contributions from the $S$-wave resonance $f_0(500)$ and $f_0(980)$ of the final state interaction, and the intermediate $P$-wave state $\bar{K}^{*}(892)$. 
In their analyzed results to the data of the CLEO Collaboration~\cite{CLEO:2011cnt}, the near-threshold enhancement in the $\pi^0\pi^0$ spectrum was attributed mainly by the $f_0(500)$ and $K^*(892)$, while the cusp-like structure around $1~\mathrm{GeV}$ was associated with the $f_0(980)$ signal. 
Furthermore, as discussed in Ref.~\cite{Ikeno:2024fjr}, even though the decay processes are similar, the $a_0(980)$ signal in the $\pi\eta$ spectra is different. 
Indeed, in Ref.~\cite{Ikeno:2024fjr}, they found that clean signal of the $a_0(980)$ in the final state interaction of the decay $D^+ \to \bar{K}^0 \pi^+ \eta$ as observed in the BESIII Collaboration's measurements~\cite{BESIII:2020pxp,BESIII:2023htx}, due to no contribution form the low-lying $\bar{K}^*$ resonance. 
This is different form the similar decay process $D^0 \to K^- \pi^+ \eta$ as investigated in the previous work~\cite{Toledo:2020zxj}, where they found that the line shape of the $a_0(980)$ signal was substantially distorted by the interference effect of the low-lying $\bar{K}^*$ resonances as observed in the Belle experiment~\cite{Belle:2020fbd}. 

With the recent measurements of the BESIII Collaboration~\cite{BESIII:2025wmd,BESIII:2025sea}, we investigate the decays $D^0 \rightarrow K_S^0\pi^0\eta$ and $D^0 \rightarrow K_S^0\pi^0 \pi^0$ within a unified dynamical framework of the weak interaction, where the final state interaction is treated in the coupled channel approach, which dynamically generates the scalar resonances as meson-meson molecular states. In this way, the $\pi^0\pi^0$ and $\pi^0\eta$ invariant-mass distributions can be analyzed with the same formalism, allowing us to clarify the respective roles of the resonances $f_0(500)$, $f_0(980)$, and $a_0(980)$. 
Our manuscript is organized as following. In Sec.~\ref{formalism}, we introduce the final state interaction formalism for the decays $D^0 \rightarrow K_S^0\pi^0\eta$ and $D^0 \rightarrow K_S^0\pi^0 \pi^0$. The results of combined fit for corresponding invariant mass distributions are presented in Sec.~\ref{results}. Finally, a short summary is shown in Sec.~\ref{summary}.

\section{Formalism}\label{formalism}
\begin{figure}[!htbp]
\includegraphics[scale=0.35]{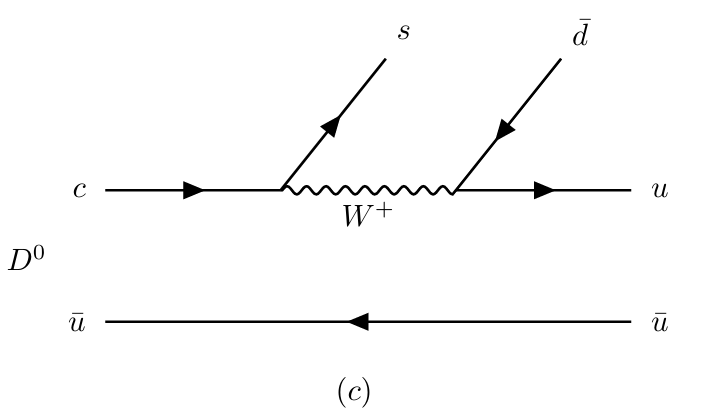}
\vspace{0.0cm} \caption{The internal $W$-emission mechanism for the decays $D^{0} \rightarrow K_S^{0} \pi^{0} \eta$ and $D^{0} \rightarrow K_S^{0} \pi^{0} \pi^0$.} \label{feyman}
\end{figure}

We investigate the $D^{0} \rightarrow K_S^{0} \pi^{0} \eta$ and $D^0 \rightarrow K_S^{0} \pi^{0} \pi^0$ reactions through the internal $W$ emission diagram, which is shown in Fig.~\ref{feyman}. The quark $c$ of the initial state $D^0$ can decay into the $s$ quark by emitting $W^+$ boson, while the quark $\bar{u}$ acts as a spectator remains unchanged. There are two ways of the hadronization. One way is that the $s \bar{d}$ quarks form the final state $\bar{K}^0$ directly, and the quark pair $u \bar{u}$ hadronizes within the pair $(u\bar{u} + d\bar{d} + s\bar{s})$ created from the vacuum. The another one is that the $u \bar{u}$ become the $\pi^0/\eta$, and the $s \bar{d}$ undergo the hadronization. 
Note that, in our cases there is no contribution from the external $W$ emission diagram, since the $\pi^0/\eta$ can not be created in this diagram, which is different from the cases of similar decays $D^+ \to \bar{K}^0 \pi^+ \eta$ and $D^0 \to K^- \pi^+ \eta$~\cite{Toledo:2020zxj,Ikeno:2024fjr}.

As done in Ref.~\cite{MartinezTorres:2009uk}, the hadronization can star from the matrix $M$ with elements $q\bar{q}$,
\begin{equation}\label{eq1}
M=\left(\begin{array}{lll}
u \bar{u} & u \bar{d} & u \bar{s} \\
d \bar{u} & d \bar{d} & d \bar{s} \\
s \bar{u} & s \bar{d} & s \bar{s}
\end{array}\right).
\end{equation}
Hence, the hadronization of the components $u \bar{u}$ and $s \bar{d}$ can be written as,
\begin{equation}\label{eq2}
\begin{aligned}
u (\bar{u}u + \bar{d}d + \bar{s}s) \bar{u}&=(M \cdot M)_{11}, \\
s (\bar{u}u + \bar{d}d + \bar{s}s) \bar{d}&=(M \cdot M)_{32}.
\end{aligned}
\end{equation}
Then, the elements of matrix $M$ can be rewritten in terms of the pseudoscalar meson fields $\phi$~\cite{Toledo:2020zxj,Bramon:1992kr,Molina:2019udw,Roca:2003uk,Gamermann:2009ouq}, 
\begin{equation}\label{eq3}
\phi =\left( 
\begin{array}{ccc}
\frac{1}{\sqrt{2}}\pi ^{0}+\frac{1}{\sqrt{3}}\eta +\frac{1}{\sqrt{6}}\eta
^{^{\prime }} & \pi ^{+} & K^{+} \\ 
\pi ^{-} & -\frac{1}{\sqrt{2}}\pi ^{0}+\frac{1}{\sqrt{3}}\eta +\frac{1}{%
\sqrt{6}}\eta ^{^{\prime }} & K^{0} \\ 
K^{-} & \bar{K}^{0} & -\frac{1}{\sqrt{3}}\eta +\frac{\sqrt{2}}{\sqrt{3}}\eta
^{^{\prime }}
\end{array}
\right). 
\end{equation}
In fact, the matrix $M$ in SU(3) should be $\phi$ + $\frac{1}{\sqrt{3}}$diag$(\eta_1, \eta_1, \eta_1)$, where the $\eta_1$ is a singlet, taking into account the standard mixing between $\eta$ and $\eta^{\prime}$. The term $\frac{1}{\sqrt{3}}$diag$(\eta_1, \eta_1, \eta_1)$ is omitted because the structure $[\phi, \partial_{\mu}\phi]$ in the chiral Lagrangians render this term inoperative. In Refs.~\cite{Liang:2014tia,Xie:2014tma}, this ordinary matrix $\phi$ of the chiral perturbation theory is also used. Then, all the possible meson-meson components in Eq.~(\ref{eq2}) are given by
\begin{equation}\label{eq4}
\begin{aligned}
\left( M\cdot M\right) _{11} &= \frac{1}{2}\pi ^{0}\pi ^{0}+\frac{1}{3}\eta \eta +\frac{2}{\sqrt{6}}\pi^{0}\eta +\pi ^{+}\pi ^{-}+K^{+}K^{-}, \\
\left( M\cdot M\right) _{32} &= K^{-}\pi ^{+}-\frac{1}{\sqrt{2}}\bar{K}^{0}\pi^{0}, 
\end{aligned}
\end{equation}
where we ignore the contributions of $\eta^{\prime}$ in our calculation as done in Ref.~\cite{Toledo:2020zxj} because it has a large mass and does not play a role in the generation of the $f_0(980)/a_0(980)$.

\begin{figure}[!htbp]
\begin{subfigure}{0.475\textwidth}
\centering
\includegraphics[width=1\linewidth]{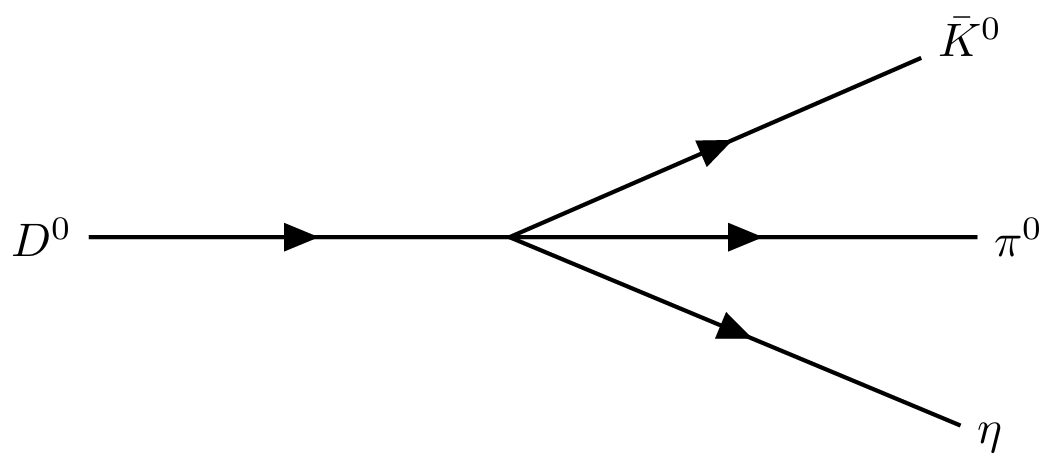} 
\caption{Tree-level production.}
\label{scatter1}
\end{subfigure}
\begin{subfigure}{0.475\textwidth}  
\centering 
\includegraphics[width=1\linewidth]{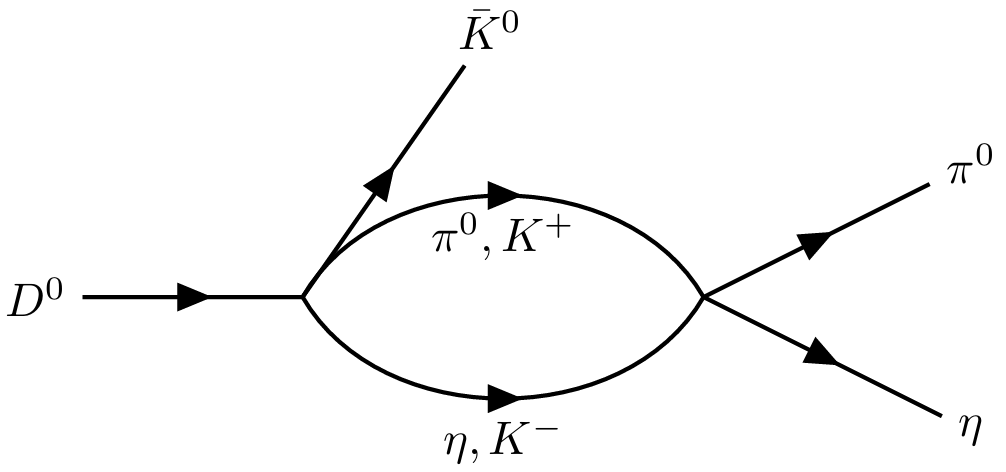} 
\caption{Rescattering of $\pi^0 \eta$ and $K^+ K^-$.}
\label{scatter2}  
\end{subfigure}	
\begin{subfigure}{0.475\textwidth}  
\centering
\includegraphics[width=1\linewidth]{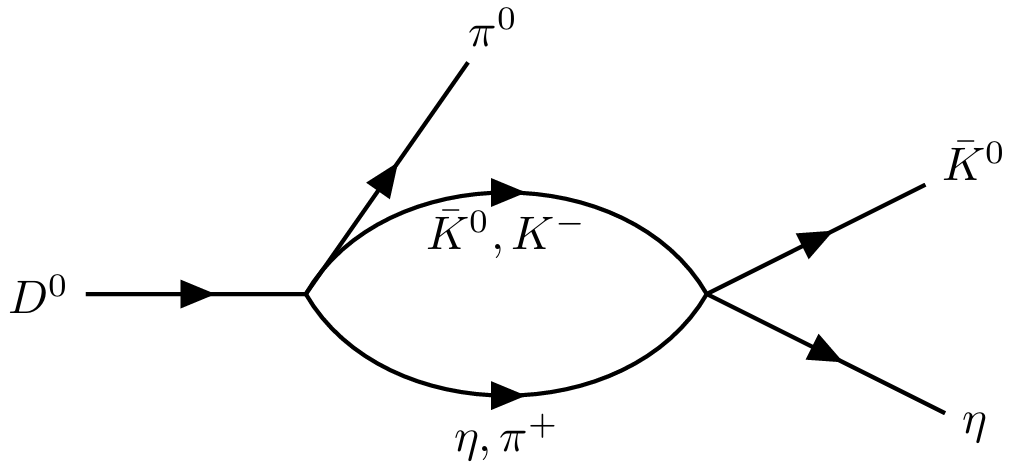} 
\caption{Rescattering of $\bar{K}^0 \eta$ and $K^- \pi^-$.}
\label{scatter3}
\end{subfigure}
\begin{subfigure}{0.475\textwidth}  
\centering
\includegraphics[width=1\linewidth]{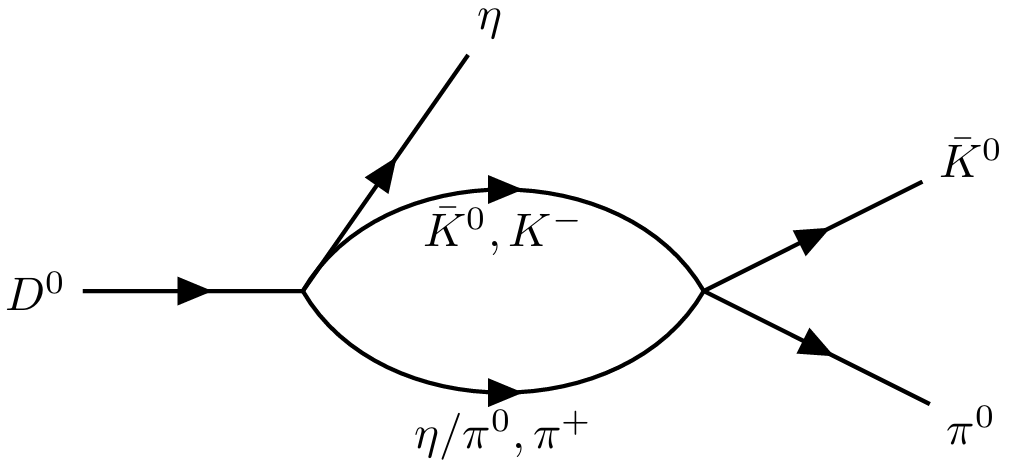} 
\caption{Rescattering of $\bar{K}^0 \eta$, $\bar{K}^0 \pi^0$, and $K^- \pi^-$.}
\label{scatter4}
\end{subfigure}
\captionsetup{justification=raggedright}
\caption{Diagrammatic representations of rescattering for the decay $D^0 \rightarrow K_S^0 \pi^0 \eta$.}
\label{scatter-Kpieta}
\end{figure}

\begin{figure}[!htbp]
\begin{subfigure}{0.475\textwidth}
\centering
\includegraphics[width=1\linewidth]{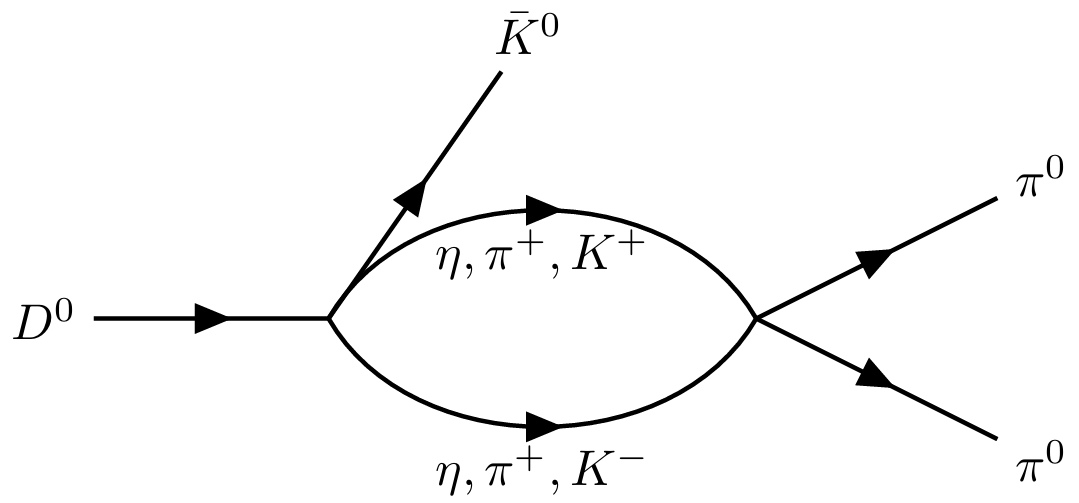} 
\caption{Rescattering of $\eta \eta$, $\pi^+ \pi^-$, and $K^+ K^-$.}
\label{scatter1-kpp}
\end{subfigure}
\begin{subfigure}{0.475\textwidth}  
\centering 
\includegraphics[width=1\linewidth]{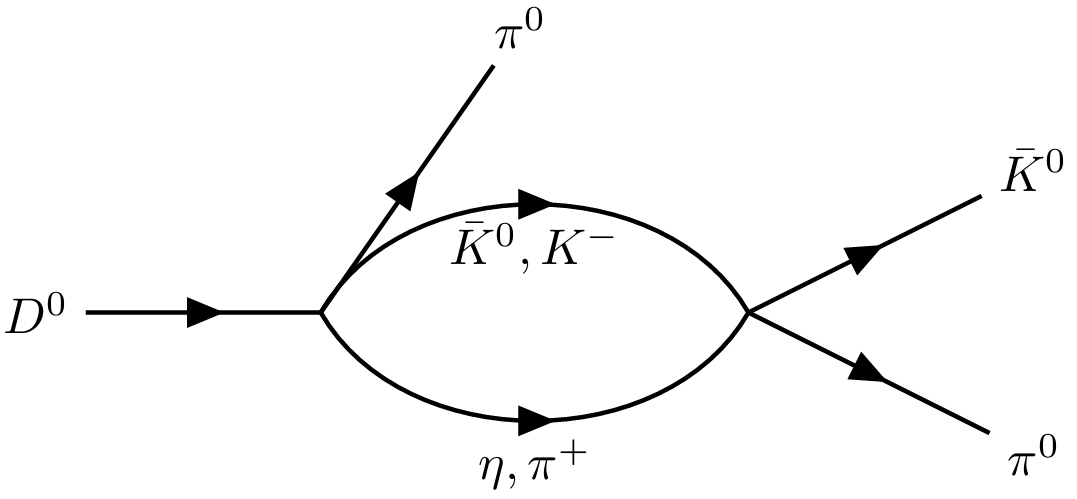} 
\caption{Rescattering of $\bar{K}^0 \eta$ and $K^- \pi^+$.}
\label{scatter2-kpp}  
\end{subfigure}	
\captionsetup{justification=raggedright}
\caption{Diagrammatic representations of rescattering for the decay $D^0 \rightarrow K_S^0 \pi^0 \pi^0$.}
\label{scatter-Kpipi}
\end{figure}

Therefore, the hadronization processes for the decays $D^{0} \rightarrow K_S^{0} \pi^{0} \eta$ and $D^0 \rightarrow K_S^{0} \pi^{0} \pi^0$ as shown in Fig.~\ref{feyman} can be expressed as
\begin{equation}\label{eq5}
\begin{aligned}
H &= V_{p}V_{cs}V_{ud}\left[ 
\left( s\bar{d}\rightarrow \bar{K}^{0}\right) \left[ M_{11}\rightarrow
\left( M\cdot M\right) _{11}\right] +\left( u\bar{u}\rightarrow \frac{1}{\sqrt{2}}\pi ^{0}\right) \left[ M_{32}\rightarrow \left( M\cdot M\right) _{32}\right] \right. \\
& \left. + \left( u\bar{u}\rightarrow \frac{1}{\sqrt{3}}\eta \right) \left[
M_{32}\rightarrow \left( M\cdot M\right) _{32} \right] \right]  \\
&= V_{p}V_{cs}V_{ud}\left[ \frac{1}{3}\bar{K}^{0}\eta \eta + \frac{1}{\sqrt{6}}\bar{K}^{0}\pi ^{0}\eta +\pi ^{+}\pi ^{-}\bar{K}^{0}+K^{+}K^{-}\bar{K}^{0} +\frac{1}{\sqrt{2}}\pi ^{0}K^{-}\pi ^{+} +\frac{1}{\sqrt{3}}\eta K^{-}\pi ^{+} \right] \\
&= C_1 \left[ \frac{1}{3}\bar{K}^{0}\eta \eta + \frac{1}{\sqrt{6}}\bar{K}^{0}\pi ^{0}\eta +\pi ^{+}\pi ^{-}\bar{K}^{0}+K^{+}K^{-}\bar{K}^{0} +\frac{1}{\sqrt{2}}\pi ^{0}K^{-}\pi ^{+} +\frac{1}{\sqrt{3}}\eta K^{-}\pi ^{+} \right],
\end{aligned}
\end{equation}
where the factor $C_1=V_{p}V_{cs}V_{ud}$ is a global parameters, which absorbs the production vertex $V_p$ and the elements of CKM matrix $V_{cs}V_{ud}$, and can be obtained by fitting experimental data as done in our former works~\cite{Wang:2021ews,Ahmed:2020qkv,Wang:2020pem,Liang:2023ekj}. Note that $C_1$ also absorbs the normalization factor when fitting the experimental data in events. 
It is obvious that the final states $\bar{K}^{0} \pi^0 \eta$ not only can produce from hadronization processes directly, but also can generate from the rescattering procedure of the final state interactions with the other terms, while the ones $\bar{K}^{0} \pi^0 \pi^0$ are only can obtain by rescattering effect, which are shown in Figs.~\ref{scatter-Kpieta} and \ref{scatter-Kpipi}, respectively. Therefore, considering the same $W$-internal emission mechanism of Fig.~\ref{feyman}, the amplitudes of the $D^0 \rightarrow K^0_S \pi^0 \eta$ and $D^0 \rightarrow K^0_S \pi^0 \pi^0$ decay processes in the $S$ wave can be written as
\begin{equation}\label{eq6}
\begin{aligned}
t\left( s_{12},s_{23}\right) _{D^{0}\rightarrow \bar{K}^{0}\pi^{0}\eta}
&=\frac{1}{\sqrt{6}}C_{1} + \frac{1}{\sqrt{6}}C_{1}G_{\pi^0 \eta }\left( s_{23}\right) T_{\pi^0 \eta \rightarrow \pi^{0}\eta}\left( s_{23}\right) +C_{1}G_{K^{+}K^{-}}\left( s_{23}\right) T_{K^{+}K^{-}\rightarrow \pi ^{0}\eta}\left( s_{23}\right) \\
&+ \frac{1}{\sqrt{6}}C_{1}G_{\bar{K}^{0}\eta }\left( s_{23}\right) T_{\bar{K}^{0}\eta \rightarrow \bar{K}^{0}\eta}\left( s_{13}\right) +\frac{1}{\sqrt{2}}C_{1}G_{K^{-}\pi ^{+}}\left( s_{13}\right) T_{K^{-}\pi^{+}\rightarrow \bar{K}^{0}\eta}\left( s_{13}\right) \\
&+ \frac{1}{3}C_{1}G_{\bar{K}^{0} \eta }\left( s_{12}\right) T_{\bar{K}^{0} \eta \rightarrow \bar{K}^{0}\pi^{0}}\left( s_{12}\right) + \frac{1}{\sqrt{6}}C_{1}G_{\bar{K}^{0} \pi^0}\left( s_{12}\right) T_{\bar{K}^{0} \pi^0 \rightarrow \bar{K}^{0} \pi^{0}}\left( s_{12}\right) \\
&+ \frac{1}{\sqrt{3}}C_{1}G_{K^{-} \pi^+}\left( s_{12}\right) T_{K^{-} \pi^+ \rightarrow \bar{K}^{0} \pi^{0}}\left( s_{12}\right), 
\end{aligned} 
\end{equation}%
\begin{equation}\label{eq7}
\begin{aligned}
t\left( s_{12},s_{23}\right) _{D^{0}\rightarrow \bar{K}^{0}\pi ^{0}\pi ^{0}}
&=\frac{1}{2} \times \frac{1}{3}C_{1}G_{\eta \eta }\left( s_{23}\right) T_{\eta \eta
\rightarrow \pi ^{0}\pi ^{0}}\left( s_{23}\right) +C_{1}G_{\pi ^{+}\pi
^{-}}\left( s_{23}\right) T_{\pi ^{+}\pi ^{-}\rightarrow \pi ^{0}\pi
^{0}}\left( s_{23}\right) \\
&+C_{1}G_{K^{+}K^{-}}\left( s_{23}\right)
T_{K^{+}K^{-}\rightarrow \pi ^{0}\pi ^{0}}\left( s_{23}\right) +\frac{1}{\sqrt{6}}C_{1}G_{\bar{K}^{0}\eta }\left( s_{12}\right) T_{\bar{K}^{0}\eta \rightarrow \bar{K}^{0}\pi ^{0}}\left( s_{12}\right)\\
& +\frac{1}{
\sqrt{2}}C_{1}G_{K^{-}\pi ^{+}}\left( s_{12}\right) T_{K^{-}\pi
^{+}\rightarrow \bar{K}^{0}\pi ^{0}}\left( s_{12}\right). 
\end{aligned} 
\end{equation}%
The $s_{ij}$ is the energy of two-body system, which is defined as $s_{ij}=(p_i+p_j)^2$, with $p_i$ and $p_j$ are the four-momenta of these two particles. The indices $i, j = 1, 2, 3$ represent three final states of $K_S^0$, $\pi^0$, and $\eta/\pi^0$, respectively. Note that there is an extra factor 1/2 for the loop function of the system $\eta \eta$ with identical particles in Eq.~(\ref{eq7}), as done in Refs.~\cite{Xie:2014tma,Zhang:2024myn}. Besides, we take $ \vert K_S^0 \rangle = \frac{1}{\sqrt{2}}( \vert K^0 \rangle - \vert \bar{K}^0 \rangle )$ into account as done in Ref.~\cite{Dai:2021owu}, and change the final state from $\bar{K}^0$ to $K^0_S$, the $S$-wave amplitude in Eqs.~(\ref{eq6}) and (\ref{eq7}) can be revised as,
\begin{equation}\label{eq8}
\begin{aligned}
t\left( s_{12},s_{23}\right) _{D^{0}\rightarrow K_S^{0}\pi^{0}\eta}
&=\frac{-1}{2\sqrt{3}}C_{1} - \frac{1}{2\sqrt{3}}C_{1}G_{\pi^0 \eta }\left( s_{23}\right) T_{\pi^0 \eta \rightarrow \pi^{0}\eta}\left( s_{23}\right) - \frac{1}{\sqrt{2}} C_{1}G_{K^{+}K^{-}}\left( s_{23}\right) T_{K^{+}K^{-}\rightarrow \pi ^{0}\eta}\left( s_{23}\right) \\
&+ \frac{1}{2\sqrt{6}}C_{1}G_{\bar{K}^{0}\eta }\left( s_{23}\right) T_{\bar{K}^{0}\eta \rightarrow \bar{K}^{0}\eta}\left( s_{13}\right) -\frac{1}{2}C_{1}G_{K^{-}\pi ^{+}}\left( s_{13}\right) T_{K^{-}\pi^{+}\rightarrow \bar{K}^{0}\eta}\left( s_{13}\right) \\
&+ \frac{1}{6}C_{1}G_{\bar{K}^{0} \eta }\left( s_{12}\right) T_{\bar{K}^{0} \eta \rightarrow \bar{K}^{0}\pi^{0}}\left( s_{12}\right) + \frac{1}{2\sqrt{6}}C_{1}G_{\bar{K}^{0} \pi^0}\left( s_{12}\right) T_{\bar{K}^{0} \pi^0 \rightarrow \bar{K}^{0} \pi^{0}}\left( s_{12}\right) \\
&- \frac{1}{\sqrt{6}}C_{1}G_{K^{-} \pi^+}\left( s_{12}\right) T_{K^{-} \pi^+ \rightarrow \bar{K}^{0} \pi^{0}}\left( s_{12}\right), 
\end{aligned} 
\end{equation}%
\begin{equation}\label{eq9}
\begin{aligned} 
t\left( s_{12},s_{23}\right) _{D^{0}\rightarrow K_S^{0}\pi ^{0}\pi ^{0}}
&=  \frac{-1}{6\sqrt{2}} C_{1}G_{\eta \eta }\left(s_{23}\right) T_{\eta \eta \rightarrow \pi ^{0}\pi ^{0}}\left( s_{23}\right) - \frac{1}{\sqrt{2}} C_{1}G_{\pi ^{+}\pi ^{-}}\left(s_{23}\right) T_{\pi ^{+}\pi ^{-}\rightarrow \pi ^{0}\pi ^{0}}\left(s_{23}\right) \\
& - \frac{1}{\sqrt{2}} C_{1}G_{K^{+}K^{-}}\left(s_{23}\right) T_{K^{+}K^{-}\rightarrow \pi ^{0}\pi ^{0}}\left( s_{23}\right) + \frac{1}{2\sqrt{6}}C_{1}G_{\bar{K}^{0}\eta }\left(s_{23}\right) T_{\bar{K}^{0}\eta \rightarrow \bar{K}^{0}\pi ^{0}}\left(s_{13}\right) \\
& - \frac{1}{2} C_{1}G_{K^{-}\pi ^{+}}\left( s_{13}\right) T_{K^{-}\pi ^{+}\rightarrow \bar{K}^{0}\pi ^{0}}\left( s_{13}\right),
\end{aligned} 
\end{equation}
where $G_{PP^{\prime}}$ and $T_{PP^{\prime}}$ are the loop functions and the two-body scattering amplitudes, respectively, see some more details below. 
Note that, for the decay $D^{0}\rightarrow K_S^{0}\pi^{0}\eta$, the $\bar{K}^0 \eta$ rescattering is disregarded as done in Ref.~\cite{Ikeno:2024fjr} for the process  $D^+ \to \bar{K}^0 \pi^+ \eta$ due to its threshold far way from energy region of the $\kappa / K^*_0(700)$.

The scattering amplitudes in Eqs.~(\ref{eq8}) and (\ref{eq9}) can be calculated by solving the coupled channel Bethe-Salpeter equation~\cite{Oset:1997it,Oller:1998zr},
\begin{equation}\label{eq10}
T = \left[ 1 - VG \right]^{-1}V,
\end{equation}
where the matrix $V$ is composed by the scattering potentials of corresponding coupled channel. In the chiral unitary approach, the interaction potentials can be evaluated from the chiral Lagrangians. For the isospin $I=0$ sector, $V$ is a $5 \times 5$ matrix of the interaction kernel for five channels: $\pi^+ \pi^-(1)$, $\pi^0 \pi^0(2)$, $K^+ K^-(3)$, $K^0 \bar{K}^0(4)$, and $\eta\eta(5)$, and the explicit expressions for these elements are given by~\cite{Wang:2021ews,Ahmed:2020qkv,Gamermann:2006nm,Liang:2014tia}
\begin{equation}\label{eq11}
\begin{aligned} 
V_{11} &=-\frac{1}{2 f^{2}} s, \quad V_{12}=-\frac{1}{\sqrt{2} f^{2}}\left(s-m_{\pi}^{2}\right), \quad V_{13}=-\frac{1}{4 f^{2}} s, \\  
V_{14} &=-\frac{1}{4 f^{2}} s, \quad V_{15}=-\frac{1}{3 \sqrt{2} f^{2}} m_{\pi}^{2}, \quad V_{22}=-\frac{1}{2 f^{2}} m_{\pi}^{2}, \\  
V_{23} &=-\frac{1}{4 \sqrt{2} f^{2}} s, \quad V_{24}=-\frac{1}{4 \sqrt{2} f^{2}} s, \quad V_{25}=-\frac{1}{6 f^{2}} m_{\pi}^{2}, \\ 
V_{33} &=-\frac{1}{2 f^{2}} s, \quad V_{34}=-\frac{1}{4 f^{2}} s, \\ 
V_{35} &=-\frac{1}{12 \sqrt{2} f^{2}}\left(9 s-6 m_{\eta}^{2}-2 m_{\pi}^{2}\right), \quad V_{44}=-\frac{1}{2 f^{2}} s, \\ 
V_{45} &=-\frac{1}{12 \sqrt{2} f^{2}}\left(9 s-6 m_{\eta}^{2}-2 m_{\pi}^{2}\right), \\ 
V_{55} &=-\frac{1}{18 f^{2}}\left(16 m_{K}^{2}-7 m_{\pi}^{2}\right), 
\end{aligned} 
\end{equation}
where $f=0.093$ GeV is the pion decay constant~\cite{Oller:1997ti}, and $m_P$ are the corresponding masses of pseudoscalar mesons which are taken from PDG~\cite{ParticleDataGroup:2024cfk}. 
For the $I=1/2$ sector, three channels are coupled, $K^{-} \pi^{+}(1)$, $\bar{K}^{0} \pi^{0}(2)$, $\bar{K}^{0} \eta(3)$, and then, the elements of $3 \times 3$ symmetric matrix are given by~\cite{Wang:2021ews,Guo:2005wp,Toledo:2020zxj} 
\begin{equation}\label{eq12}
\begin{aligned} 
V_{11}&= \frac{-1}{6 f^{2}}\left[\frac{3}{2} s-\frac{3}{2 s}\left(m_{\pi}^{2}-m_{K}^{2}\right)^{2}\right], \\ 
V_{12}&= \frac{1}{2 \sqrt{2} f^{2}}\left[\frac{3}{2} s-m_{\pi}^{2}-m_{K}^{2}-\frac{\left(m_{\pi}^{2}-m_{K}^{2}\right)^{2}}{2 s}\right], \\ 
V_{13}&= \frac{1}{2 \sqrt{6} f^{2}}\left[\frac{3}{2} s-\frac{7}{6} m_{\pi}^{2}-\frac{1}{2} m_{\eta}^{2}-\frac{1}{3} m_{K}^{2}+\frac{3}{2 s}\left(m_{\pi}^{2}-m_{K}^{2}\right)\left(m_{\eta}^{2}-m_{K}^{2}\right)\right], \\ 
V_{22}&= \frac{-1}{4 f^{2}}\left[-\frac{s}{2}+m_{\pi}^{2}+m_{K}^{2}-\frac{\left(m_{\pi}^{2}-m_{K}^{2}\right)^{2}}{2 s}\right], \\ 
V_{23}&=-\frac{1}{4 \sqrt{3} f^{2}}\left[\frac{3}{2} s-\frac{7}{6} m_{\pi}^{2}-\frac{1}{2} m_{\eta}^{2}-\frac{1}{3} m_{K}^{2} +\frac{3}{2 s}\left(m_{\pi}^{2}-m_{K}^{2}\right)\left(m_{\eta}^{2}-m_{K}^{2}\right)\right], \\ 
V_{33}&=-\frac{1}{4 f^{2}}\left[-\frac{3}{2} s-\frac{2}{3} m_{\pi}^{2}+m_{\eta}^{2}+3 m_{K}^{2}-\frac{3}{2 s}\left(m_{\eta}^{2}-m_{K}^{2}\right)^{2}\right]. 
\end{aligned} 
\end{equation}
For the $I=1$ sector, three channels are coupled, $K^+ K^-(1)$, $K^0 \bar{K}^0(2)$, $\pi^{0} \eta(3)$, and then, the elements of $3 \times 3$ symmetric matrix are given by~\cite{Ikeno:2024fjr,Lin:2021isc} 
\begin{equation}\label{eq13}
\begin{aligned}
V_{11}=-\frac{1}{2 f^2} s, \quad V_{12}=-\frac{1}{4 f^2} s, \quad V_{13}=-\frac{1}{3\sqrt{6} f^2}\left[3 s-2m_K^2-m_\eta^2\right], \\
V_{22}=-\frac{1}{2 f^2} s, \quad V_{23}=\frac{1}{3\sqrt{6} f^2}\left[3 s-2m_K^2-m_\eta^2\right], \quad V_{33}=-\frac{2}{3 f^2} m_\pi^2,
\end{aligned}
\end{equation}
Note that, the unitary normalization $\vert \eta\eta \rangle \rightarrow \frac{1}{\sqrt{2}}\vert \eta\eta \rangle$ is accounted for the identitcal particles for the $G$ function without an extra factor in Eq.~(\ref{eq10})~\cite{Liang:2014tia}. Generally, the loop function $G$ are logarithmically, we adopt three-momentum cut-off method to solve this problem, and the explicit form is given by~\cite{Guo:2005wp}
\begin{equation}\label{eq14}
\begin{aligned}
G(s)= & \frac{1}{16 \pi^2 s}\left\{\sigma\left(\arctan \frac{s+\Delta}{\sigma \lambda_1}+\arctan \frac{s-\Delta}{\sigma \lambda_2}\right)\right. \\
& \left.-\left[(s+\Delta) \ln \frac{\left(1+\lambda_1\right) q_{\max }}{m_1}+(s-\Delta) \ln \frac{\left(1+\lambda_2\right) q_{\max }}{m_2}\right]\right\},
\end{aligned}
\end{equation}
where $\sigma=[-(s-(m_1+m_2)^2)(s-(m_1-m_2)^2)]^{1/2}$, $\Delta=m_1^2-m_2^2$, and $\lambda_i=\sqrt{1+m_i^2/q^2_{max}}(i=1,2)$. 
$q_{max}$ is the cut-off momentum, and we take the value as 650 MeV, which also used in Refs.~\cite{Lin:2021isc} for the case of considering the $\eta-\eta^\prime$ mixing.

Besides, in the chiral unitary approach, the most reliable range of the $S$-wave pseudoscalar meson-pseudoscalar meson interactions is up to $1.1 \sim 1.2$GeV. In order to make reasonable calculations, we smoothly extrapolate $G(s)T(s)$ above the energy cut $\sqrt{s} \geq \sqrt{s_{cut}}=1.1$GeV, as done in Ref.~\cite{Debastiani:2016ayp},
\begin{equation}\label{eq15}
G(s) T(s)=G\left(s_{\mathrm{cut}}\right) T\left(s_{\mathrm{cut}}\right) e^{-\alpha\left(\sqrt{s}-\sqrt{s_{\mathrm{cut}}}\right)}, \quad  \sqrt{s}>\sqrt{s_{\mathrm{cut}}},
\end{equation}
where $G$ and $T$ are the loop functions and amplitudes which are introduced in Eqs.~(\ref{eq14}) and (\ref{eq10}), respectively. The $\alpha$ is the smoothing extrapolation factor, of which the value is obtained by fitting experimental data.

\begin{figure}[!htbp]
\includegraphics[scale=0.30]{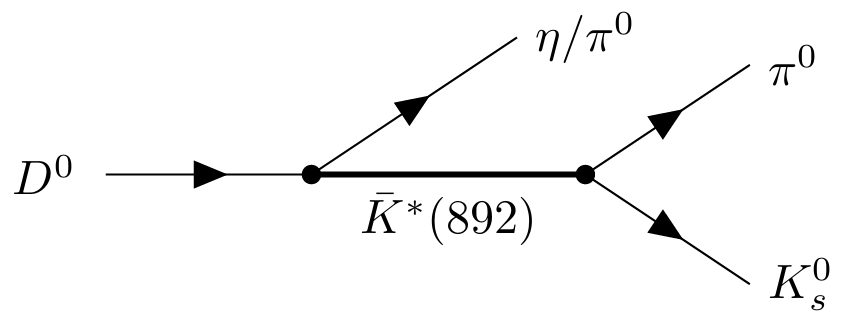}
\vspace{0.0cm} \caption{Mechanism for the $D^{0} \rightarrow K_S^{0} \pi^{0} \eta$ and $D^{0} \rightarrow K_S^{0} \pi^{0} \pi^0$ decays via the intermediate vector meson $\bar{K}^*(892)$.} \label{intermediate}
\end{figure}

In addition to the contributions from the $S$-wave pseudoscalar meson-pseudoscalar menson interactions, we also considered the contribution from the intermediate vector meson $\bar{K}^*(892)$. The mechanism is depicted in Fig.~\ref{intermediate}, and the amplitude for the process $D^0 \rightarrow \bar{K}^*(892) \pi^0 \rightarrow \bar{K}^0 \pi^0 \pi^0$ can be calculated by~\cite{Toledo:2020zxj,Roca:2020lyi},
\begin{equation}\label{eq16}
\begin{aligned}
\mathcal{M}_{\bar{K}^*(892) }\left(s_{12}, s_{23}\right) & =  \frac{\mathcal{D} e^{i \alpha_{\bar{K}^*(892)}}}{s_{12}-M_{\bar{K}^*(892)}^2+i M_{\bar{K}^*(892)} \Gamma_{\bar{K}^*(892)^{+}}} \\
& \times\left[\frac{\left(m_{D^{0}}^2-m_{\eta/\pi^0}^2\right)\left(m_{K^0}^2-m_{\pi^{0}}^2\right)}{M_{\bar{K}^*(892)}^2}-s_{13}+s_{23}\right],
\end{aligned}
\end{equation}
where $\mathit{D}$ and $\alpha_{\bar{K}^*}(892)$ are the normalization factor and phase, respectively, which can be determined by fitting the experimental data. $M_{\bar{K}^*(892)}$ and $\Gamma_{\bar{K}^*(892)}$ are the mass and width of the intermediate state $\bar{K}^*(892)$, of which the values are taken from the PDG as $0.89167$ and $0.0514$ GeV~\cite{ParticleDataGroup:2024cfk}. Note that the $s_{ij}$ are not independent and fulfill following constraint condition,
\begin{equation}\label{eq17}
s_{12} + s_{13} + s_{23} = m_{D^0}^2 + m_{\bar{K}^0}^2 + m_{\pi^0}^2 + m_{\eta/\pi^0}^2.
\end{equation}
Hence, there are only two $s_{ij}$ are independent. The total amplitudes of the $D^0 \rightarrow K_S^0 \pi^0 \eta$ and $D^0 \rightarrow K_S^0 \pi^0 \pi^0$ processes which contain the contributions form the $S$ and $P$ wave can be written as,
\begin{equation}\label{eq18}
t^{\prime}(s_{12},s_{23})_{D^0 \rightarrow K_S^0 \pi^0 \eta} = t(s_{12},s_{23})_{D^0\rightarrow K_S^0 \pi^0 \eta}+\mathcal{M}_{\bar{K}^*(892) }\left(s_{12}, s_{23}\right),
\end{equation}
\begin{equation}\label{eq19}
t^{\prime}(s_{12},s_{23})_{D^0 \rightarrow K_S^0 \pi^0 \pi^0} = t(s_{12},s_{23})_{D^0\rightarrow K_S^0 \pi^0 \pi^0}+\mathcal{M}_{\bar{K}^*(892) }\left(s_{12}, s_{23}\right) + (2 \leftrightarrow 3).
\end{equation}
Note that, when considering the identical particles in $D^0\rightarrow K_S^0 \pi^0 \pi^0$ decay, the parameters in the contributions of the $P$ wave are different, there are four parameters, and thus, two normalizations $D/D_{1}$ and two phases $\alpha_{\bar{K}^*(892)}/\alpha^{\prime}_{\bar{K}^*(892)}$.

Finally, the double differential width distribution of these two three-body decays are written as,
\begin{equation}\label{eq20}
\frac{d^2 \Gamma}{d s_{12} d s_{23}}=\frac{1}{(2 \pi)^3} \frac{1}{32 m_{D^{0}}^3}\left|t^{\prime}\left(s_{12}, s_{23}\right)_{D^0 \rightarrow K_S^0 \pi^0 \eta}\right|^2,
\end{equation}
\begin{equation}\label{eq21}
\frac{d^2 \Gamma}{d s_{12} d s_{23}}=\frac{1}{(2 \pi)^3} \frac{1}{32 m_{D^{0}}^3}\frac{1}{2}\left|t^{\prime}\left(s_{12}, s_{23}\right)_{D^0\rightarrow K_S^0 \pi^0 \pi^0}\right|^2,
\end{equation}
where the factor $\frac{1}{2}$ origins from the identical particle $\pi^0$ in final states, and the $t^{\prime}(s_{12},s_{23})$ is the total amplitudes which are presented in Eqs.~(\ref{eq18}) and (\ref{eq19}). With the constraint condition given by Eq.~(\ref{eq17}), one can calculate the invariant mass distributions $d\Gamma/ds_{12}$, $d\Gamma/ds_{23}$ and $d\Gamma/ds_{23}$ by integrating over the other variable with the limit of Dalitz plot, where more detail can be found in PDG~\cite{ParticleDataGroup:2024cfk}.

\section{results and discussions}\label{results}
In our formalism discussed above, there are several parameters need to determined by fitting the experimental data. For the $D^0\rightarrow K_S^0 \pi^0 \eta$ process, there are four parameters, the global parameter $C_1$, the extrapolation parameter $\alpha$, the normalization parameters $D$ and  phases $\alpha_{\bar{K}^*(892)}$ in the $P$-wave amplitude. For the $D^0\rightarrow K_S^0 \pi^0 \pi^0$ decay, due to the identical particles, there are two more parameters for the $P$ wave, $D_{1}$ and $\alpha^{\prime}_{\bar{K}^*(892)}$. As shown in Eq.~(\ref{eq18}) and (\ref{eq19}), the coherence of the amplitudes between the $S$ and $P$ waves are taken into account.

\begin{figure}[!htbp]
\begin{subfigure}{0.52\textwidth}
\centering
\includegraphics[width=1\linewidth]{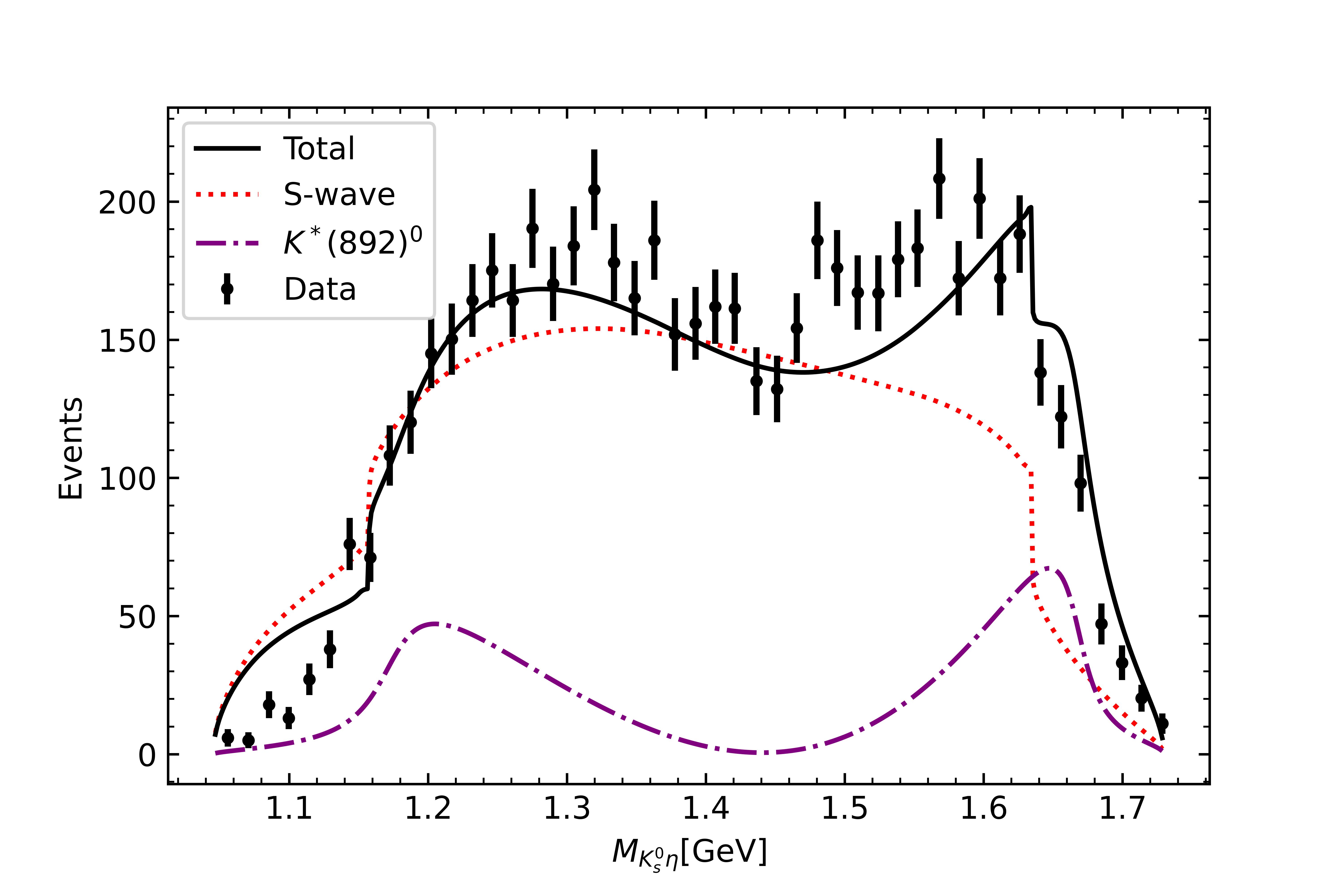} 
\caption{Invariant mass distribution of $K_S^0 \eta$.}
\label{Contribution1}
\end{subfigure}
\begin{subfigure}{0.475\textwidth}  
\centering 
\includegraphics[width=1\linewidth]{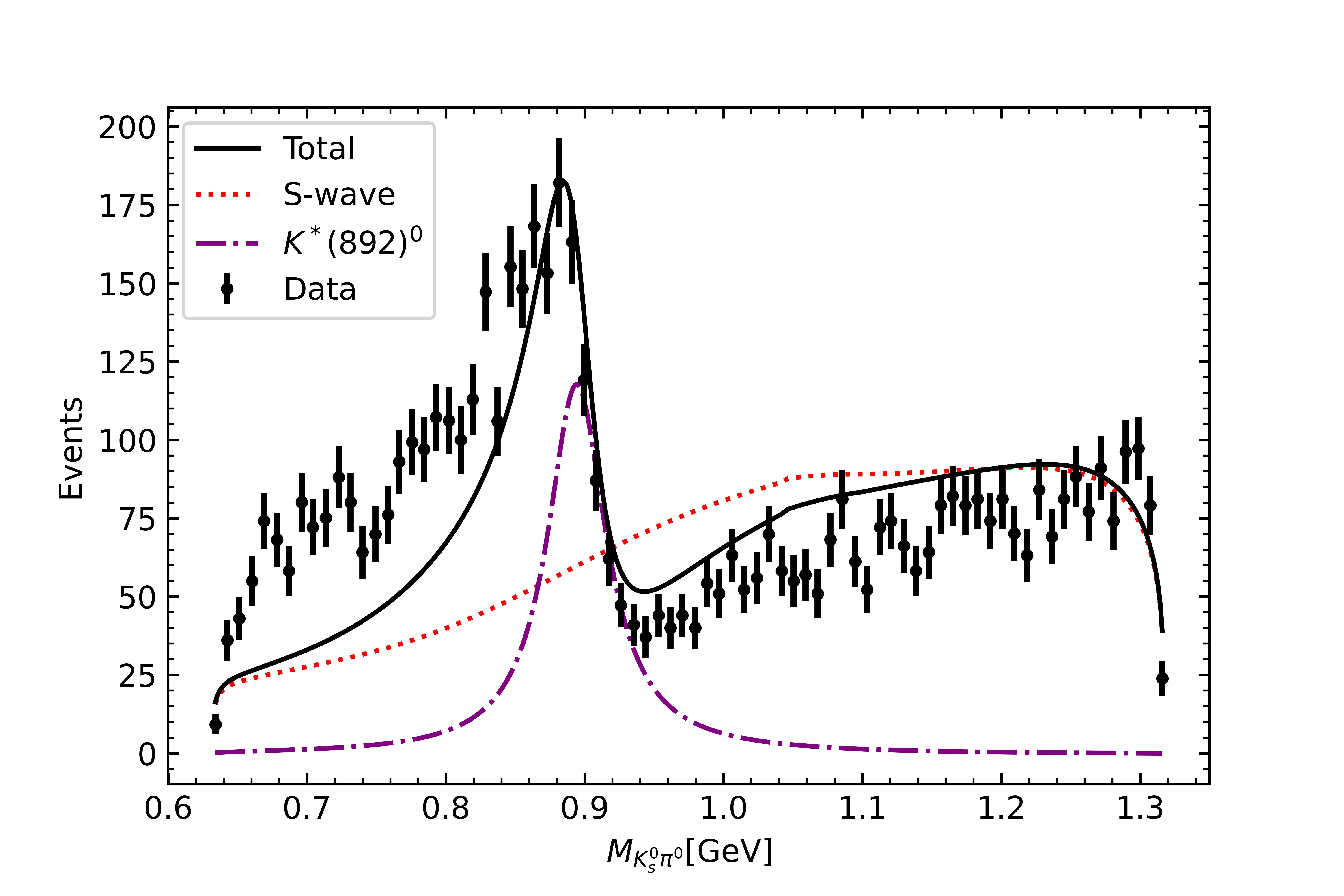} 
\caption{Invariant mass distribution of $K_S^0 \pi^0$.}
\label{Contribution2}  
\end{subfigure}	
\begin{subfigure}{0.475\textwidth}  
\centering
\includegraphics[width=1\linewidth]{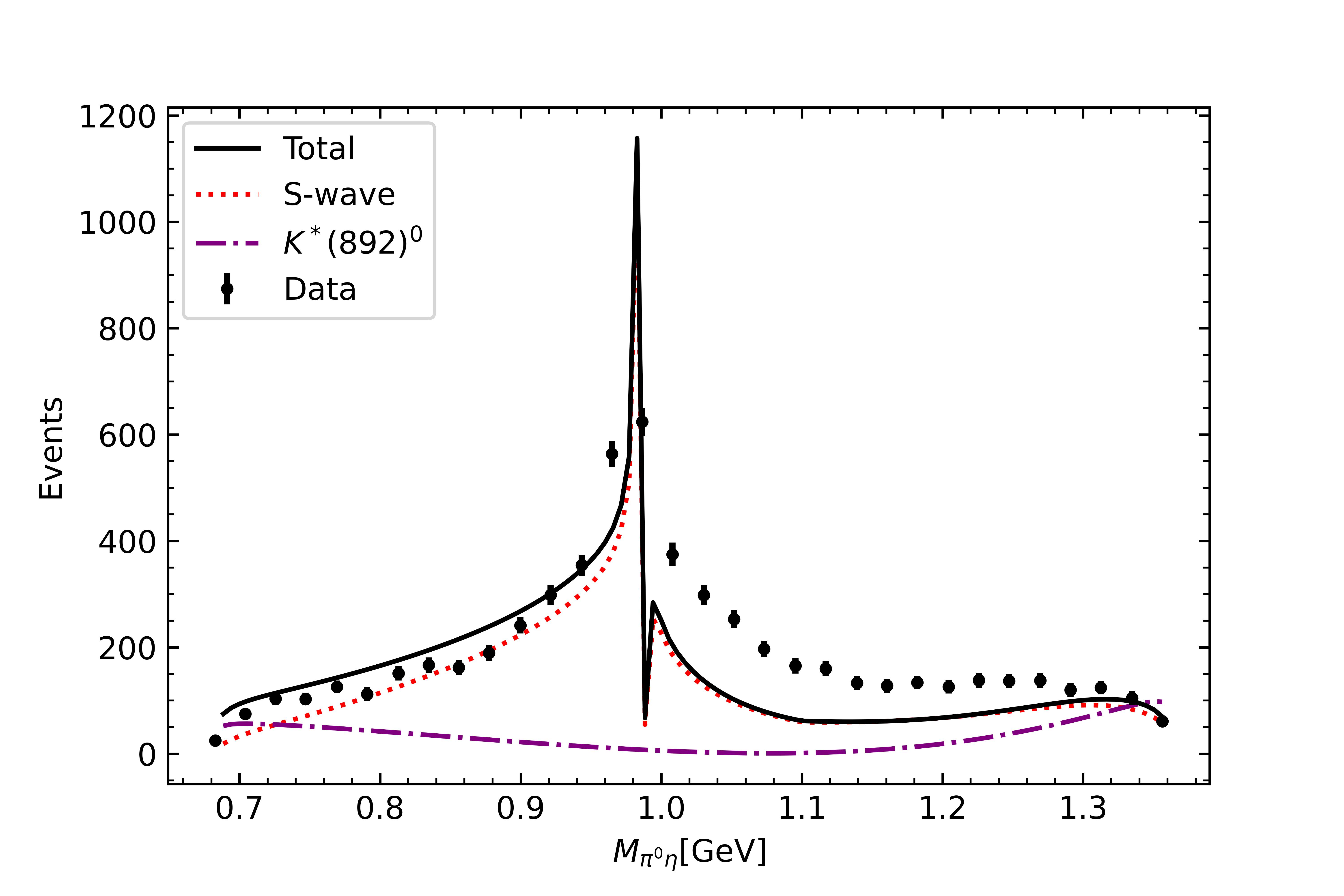} 
\caption{Invariant mass distribution of $\pi^0 \eta$.}
\label{Contribution3}
\end{subfigure}
\captionsetup{justification=raggedright}
\caption{Combined fit for the invariant mass distributions of the decay $D^0 \rightarrow K_S^0 \pi^0 \eta$. The solid (black) line is the total contributions of the $S$ and $P$ waves, the dash (red) line represents the $S$-wave contributions, the dash-dot (purple) line  means the contributions of the $\bar{K}^*(892)$ in the $P$ wave. The dot (black) points are the experimental data measured by the BESIII Collaboration~\cite{BESIII:2025wmd}.}
\label{fig-Kpieta}
\end{figure}

For the $D^0 \rightarrow K_S^0 \pi^0 \eta$ process, in order to describe the $K_S^0 \pi^0$, $K_S^0 \eta$ and $\pi^0 \eta$ invariant mass distributions simultaneously, we make a combined fit to the experimental data of the BESIII Collaboration~\cite{BESIII:2025wmd}. The fitted parameters are presented in Table~\ref{parameters-Kpieta}, and the invariant mass spectra are shown in Fig.~\ref{fig-Kpieta}, where we have taken into account the bin sizes different of the spectra, see the results of Ref.~\cite{BESIII:2025wmd}. Our fitted $\chi^2/dof.$ is a bit large, since we do not considered the contributions from the other higher resonances, as considered in the experiment~\cite{BESIII:2025wmd}. One can see that, our fitting results are almost in agreement with experiments within the uncertainties. For the $K_S^0 \eta$ mass distribution as shown in Fig.~\ref{Contribution1}, there are a double hump structures, which are the contributions of the interference effect between the $S$-wave amplitude and the $P$-wave one with the intermediate resonance $\bar{K}^*(892)$, and where the $S$-wave contribution is dominant. In Fig.~\ref{Contribution2}, the structure around 0.9 GeV in the $K_S^0 \pi^0$ spectrum is the contribution of the $\bar{K}^*(892)$, and the bump in higher region benefits from the $S$ wave.  For the $\pi^0 \eta$ invariant mass distribution as depicted in Fig.~\ref{Contribution3}, the clear enhancement around 1.0 GeV is the signal of the $a_{0}(980)$, which is dynamically generated from the $S$-wave final state interactions regarded as a $K\bar{K}$ molecular state, and the contribution from the $P$-wave is tiny.

\begin{table}[!htb]
\centering
\caption{Values of the parameters from the combined fit for the data of the $K_S^0 \pi^0$, $K_S^0 \eta$ and $\pi^0 \eta$ spectra observed by the BESIII Collaboration~\cite{BESIII:2025wmd}.} \label{parameters-Kpieta}
\resizebox{0.6\textwidth}{!}
{\begin{tabular}{cccccc}
\hline\hline
Parameters  & $C_1$ &  $\alpha$ & $D$ & $\alpha_{\bar{K}^*(892)}$  & $\chi^2/dof.$\\
\hline
Fit  &  3342.04   & -3.38  & 138.31 & 2.34 & 11.59 \\
\hline\hline
\end{tabular}}
\end{table}

\begin{figure}[!htbp]
\begin{subfigure}{0.52\textwidth}
\centering
\includegraphics[width=1\linewidth]{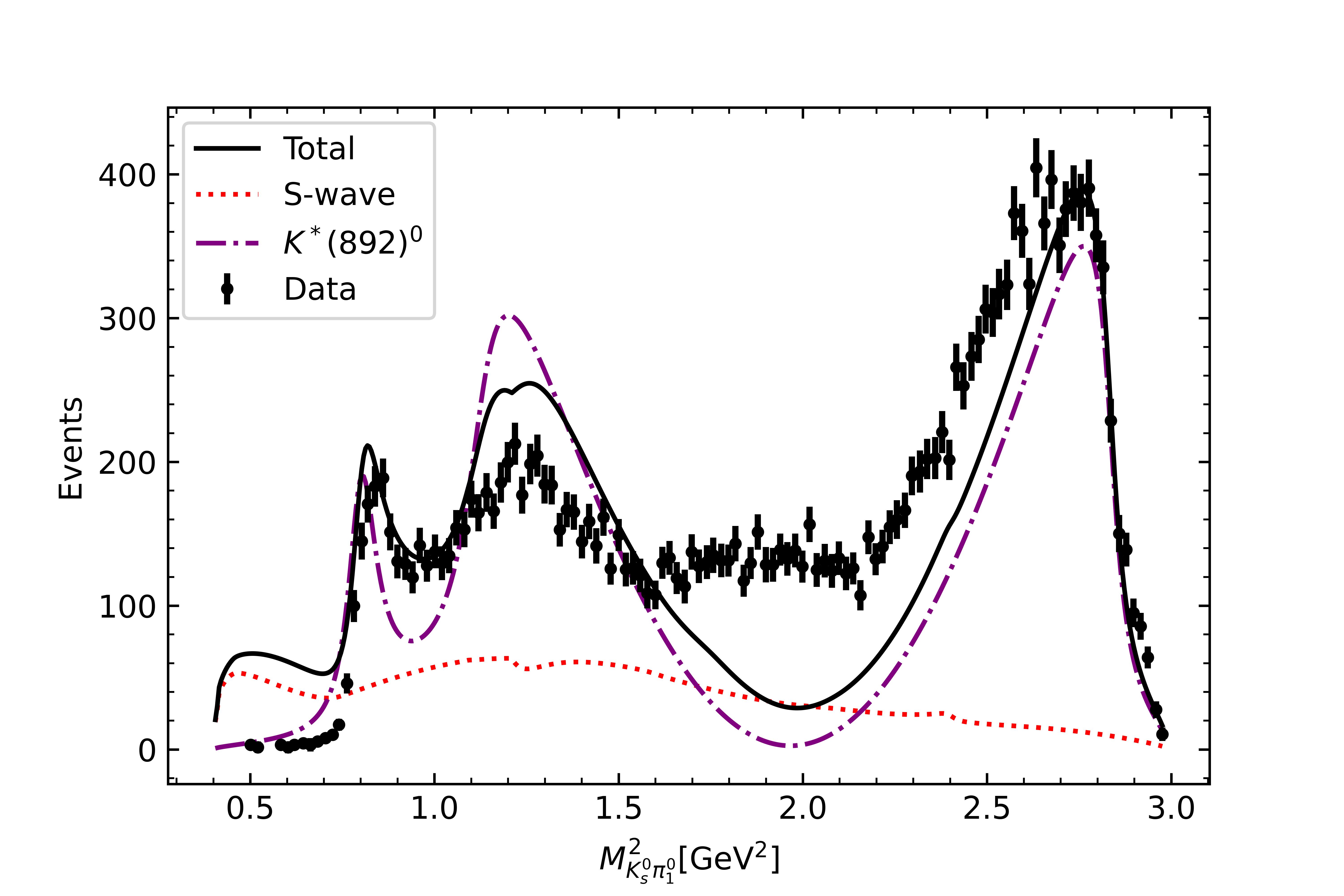} 
\caption{Invariant mass distribution of $K_S^0 \pi^0_1$.}
\label{Contribution4}
\end{subfigure}
\begin{subfigure}{0.475\textwidth}  
\centering 
\includegraphics[width=1\linewidth]{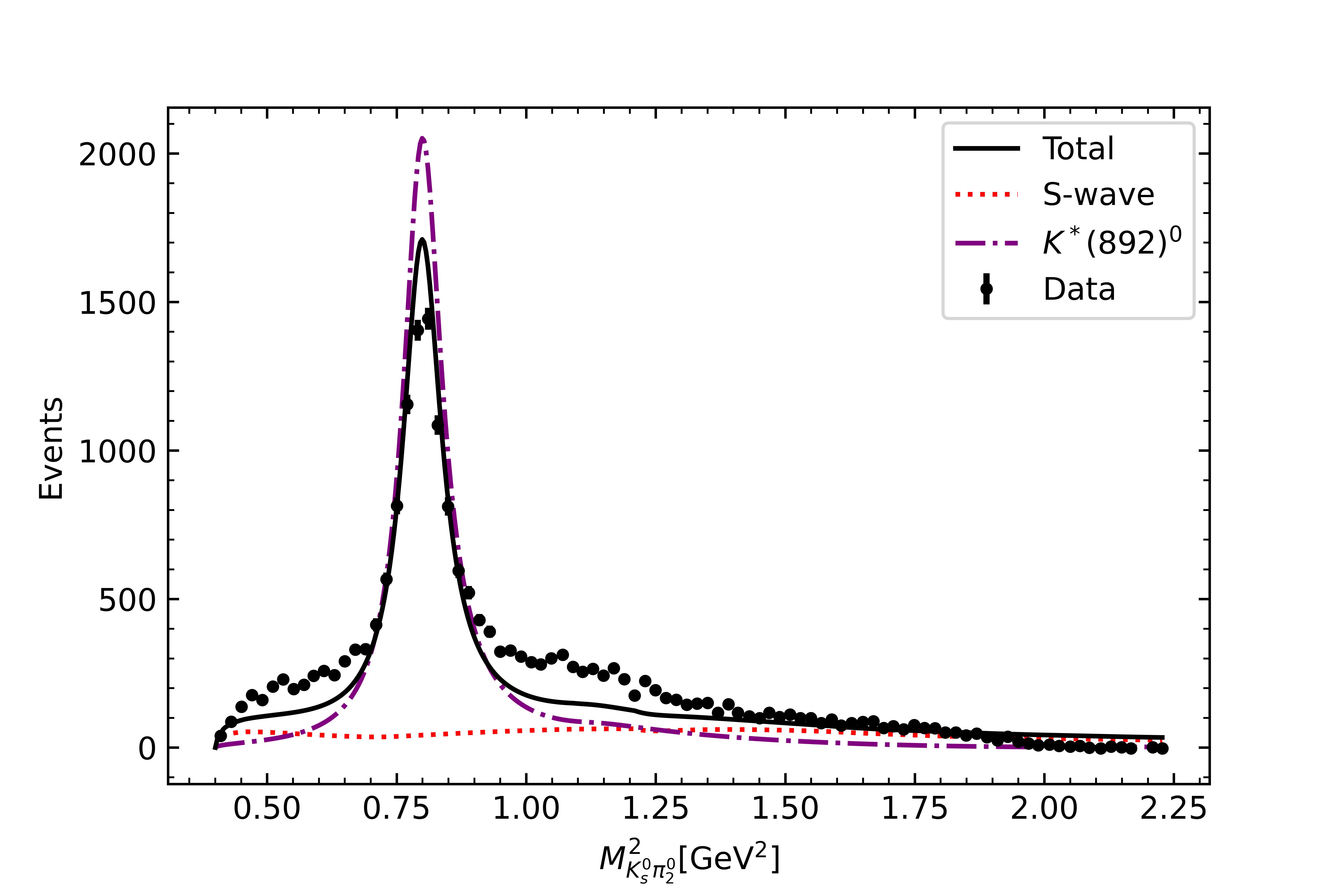} 
\caption{Invariant mass distribution of $K_S^0 \pi^0_2$.}
\label{Contribution5}  
\end{subfigure}	
\begin{subfigure}{0.475\textwidth}  
\centering
\includegraphics[width=1\linewidth]{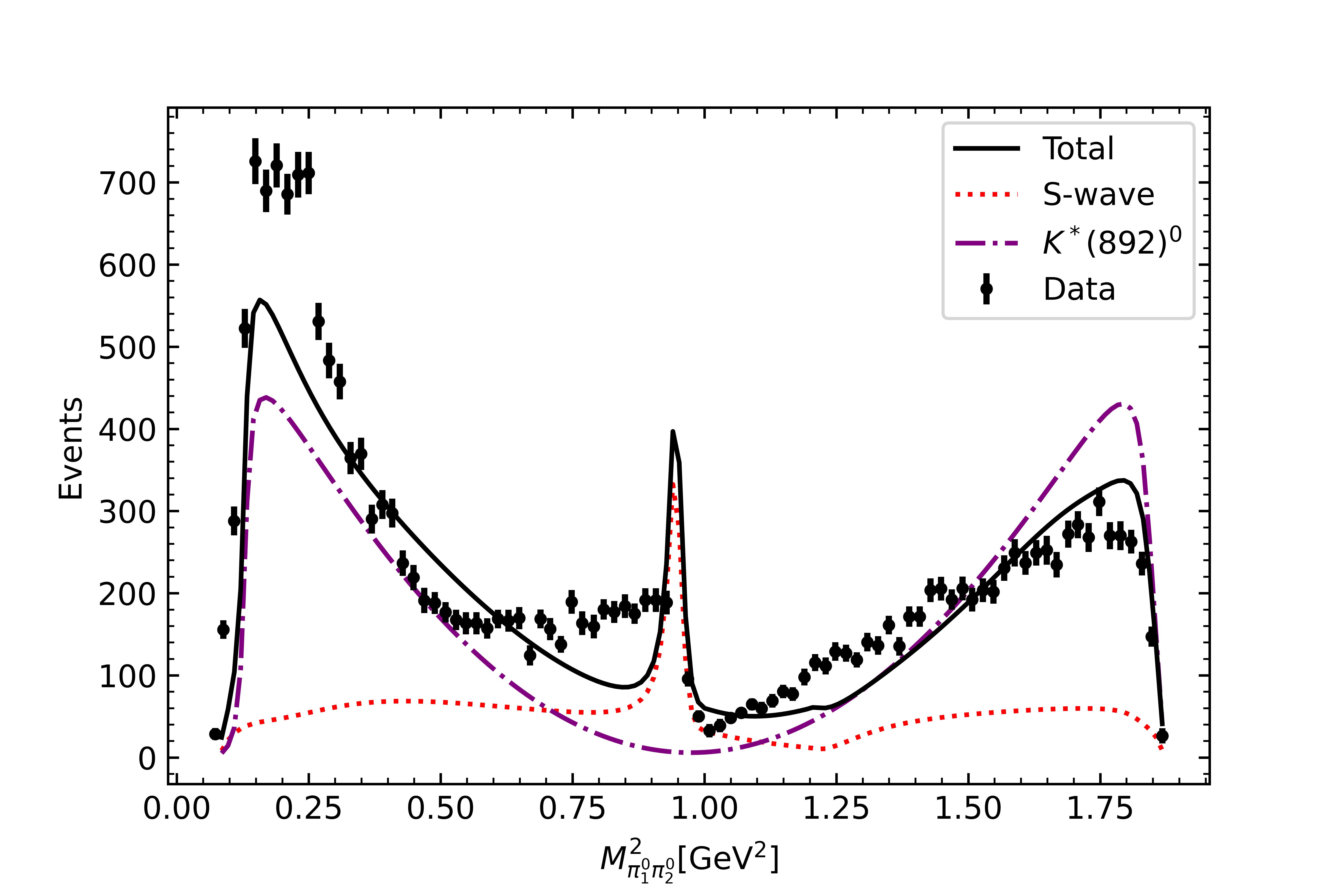} 
\caption{Invariant mass distribution of $\pi_1^0 \pi_2^0$.}
\label{Contribution6}
\end{subfigure}
\captionsetup{justification=raggedright}
\caption{Combined fit for the invariant mass distributions of the decay $D^0 \rightarrow K_S^0 \pi^0 \pi^0$. The solid (black) line is the total contributions of the $S$ and $P$ waves, the dash (red) line represents the $S$-wave contributions, the dash-dot (purple) line means the contributions of the $\bar{K}^*(892)$ in the $P$ wave. The dot (black) points are the experimental data measured by the BESIII Collaboration~\cite{BESIII:2025sea}.}
\label{fig-Kpipi}
\end{figure}

Similarly, for the $D^0 \rightarrow K_S^0 \pi^0 \pi^0$ decay, through the interference between the $S$ and $P$ waves, we also perform a combined fit to the experimental data from three corresponding invariant mass distributions measured by the BESIII Collaboration~\cite{BESIII:2025sea}, and the free parameters and mass distributions are shown in Table.~\ref{parameters-Kpipi} and Fig.~\ref{fig-Kpipi}, respectively. It looks like that the fitted $\chi^2/dof.$ is a bit big, due to the other higher resonances not considered, such as the states $\bar{K}^*_2,\ f_2$, as found in the experiment~\cite{BESIII:2025sea}. In Fig.~\ref{Contribution4}, the $K_S^0 \pi^0_1$ distribution is depicted, where three obvious structures mainly benefit from the vector meson $\bar{K}^*(892)$, and the contributions from the $S$ wave is relatively small. In Fig.~\ref{Contribution5}, our results for the $K_S^0 \pi^0_2$ mass distribution is in agreement with the BESIII measurements. The narrow peak structure around 0.75 GeV$^2$ are contributed from the vector meson $\bar{K}^*(892)$ in the $P$-wave, while the contribution from $S$-wave pseudoscalar meson-pseudoscalar interactions is just shown as a smooth background in this distribution. For the $\pi_1^0 \pi_2^0$ invariant mass distribution as presented in Fig.~\ref{Contribution6}, the structure around 0.25 GeV$^2$ is the contributions of the scalar meson $f_0(500)$ in the $S$-wave and the reflection of the vector meson $\bar{K}^*(892)$ in the $P$-wave, which are also contributed to the one in the higher energy region around 1.75 GeV$^2$. The clear peak at 1.0 GeV$^2$ is the signal of the scalar meson $f_0(980)$, which is dynamically generated from the $S$-wave pseudoscalar meson-pseudoscalar interactions. It should be mentioned that, our results are consistent with the results obtained in Ref.~\cite{Zhang:2024myn}.

\begin{table}[!htb]
\centering
\caption{Values of the parameters from the combined fit for the data of the decay $D^0 \rightarrow K_S^0 \pi^0 \pi^0$ observed by the BESIII Collaboration~\cite{BESIII:2025sea}.} \label{parameters-Kpipi}
\resizebox{0.8\textwidth}{!}
{\begin{tabular}{cccccccc}
\hline\hline
Parameters  & $C_1$ &  $\alpha$ & $D$ & $\alpha_{\bar{K}^*(892)}$ & $D_1$ & $\alpha^{\prime}_{\bar{K}^*(892)}$ & $\chi^2/dof.$\\
\hline
Fit  &  6040.13   & 31.36  & 122.10 & 1.09 & 522.41 &$1.47 \times 10^{-2}$ & 27.94\\
\hline\hline
\end{tabular}}
\end{table}

\section{Summary}\label{summary}

Motivated by the BESIII Collaboration measurements for the decays $D^0 \rightarrow K_S^0 \pi^0 \eta$ and $D^0 \rightarrow K_S^0 \pi^0 \pi^0$, we investigate these two processes with the same weak decay diagram of internal $W$ emission mechanism in the quark level. 
For the final state interaction in the hadron level, we consider the $S$-wave pseudoscalar meson-pseudoscalar meson interaction within the chiral unitary approach, where the scalar resonances $f_0(500)$, $f_0(980)$, and $a_0(980)$ are dynamically generated. 
Moreover, the contribution from the $P$-wave intermediate vector meson $\bar{K}^*(892)$ is also taken into account. 
By considering the interference of the $S$- and $P$-waves amplitudes, we make a combined fit for corresponding invariant mass spectra, where the free parameters can be obtained and gotten good descriptions of the experimental data. For the decay $D^0 \rightarrow K_S^0 \pi^0 \eta$, the peak around 1.0 GeV in the $\pi^0 \eta$ mass distribution is the signal of the $a_0(980)$, while the contribution of intermediate resonance $\bar{K}^*(982)$ in this distribution is very small and just like as a background. For the process $D^0 \rightarrow K_S^0 \pi^0 \pi^0$, the obvious structure around 1.0 GeV$^2$ in the $\pi_1^0 \pi_2^0$ mass spectrum is associated with the $f_0(980)$, which is dynamically generated from the $S$-wave coupled channel interactions, and the bump at low region benefits from the contributions of the $f_0(500)$ and $\bar{K}^*(892)$.

\section*{Acknowledgements}

We thank the useful discussions with Prof. Bai-Cian Ke.
This work is supported by the Natural Science Foundation of Guangxi province under Grant No. 2023JJA110076, the Natural Science Foundation of Hunan province under Grant No. 2023JJ30647, and the National Natural Science Foundation of China under Grants No. 12365019 and No. 12575081.


\begin{thebibliography}{99}

\bibitem{Cheng:2010cb}
H.~Y.~Cheng,
Int. J. Mod. Phys. Conf. Ser. \textbf{02}, 61-66 (2011)
[arXiv:1011.0790 [hep-ph]].

\bibitem{Cheng:2010ry}
H.~Y.~Cheng and C.~W.~Chiang,
Phys. Rev. D \textbf{81}, 074021 (2010)
[arXiv:1001.0987 [hep-ph]].

\bibitem{ParticleDataGroup:2024cfk}
S.~Navas \textit{et al.} [Particle Data Group],
Phys. Rev. D \textbf{110}, no.3, 030001 (2024).

\bibitem{Chen:2016qju}
  H.~X.~Chen, W.~Chen, X.~Liu and S.~L.~Zhu,
  Phys.\ Rept.\  {\bf 639}, 1 (2016)
  [arXiv:1601.02092 [hep-ph]].

\bibitem{Hosaka:2016pey}
  A.~Hosaka, T.~Iijima, K.~Miyabayashi, Y.~Sakai and S.~Yasui,
  PTEP {\bf 2016}, no. 6, 062C01 (2016)
  [arXiv:1603.09229 [hep-ph]].

\bibitem{Guo:2017jvc}
  F.~K.~Guo, C.~Hanhart, U.-G.~Mei{\ss}ner, Q.~Wang, Q.~Zhao and B.~S.~Zou,
  Rev.\ Mod.\ Phys.\  {\bf 90}, no. 1, 015004 (2018)
  [arXiv:1705.00141 [hep-ph]].

\bibitem{Esposito:2016noz}
  A.~Esposito, A.~Pilloni and A.~D.~Polosa,
  Phys.\ Rept.\  {\bf 668}, 1 (2016)
  [arXiv:1611.07920 [hep-ph]].

\bibitem{Olsen:2017bmm}
  S.~L.~Olsen, T.~Skwarnicki and D.~Zieminska,
  Rev.\ Mod.\ Phys.\  {\bf 90}, no. 1, 015003 (2018)
  [arXiv:1708.04012 [hep-ph]].
  
\bibitem{Yuan:2018inv}
  C.~Z.~Yuan,
  Int.\ J.\ Mod.\ Phys.\ A {\bf 33}, no. 21, 1830018 (2018)
  [arXiv:1808.01570 [hep-ex]].

\bibitem{Brambilla:2019esw}
N.~Brambilla, S.~Eidelman, C.~Hanhart, A.~Nefediev, C.~P.~Shen, C.~E.~Thomas, A.~Vairo and C.~Z.~Yuan,
Phys. Rept. \textbf{873}, 1-154 (2020)
[arXiv:1907.07583 [hep-ex]].

\bibitem{Wang:2021ail}
W.~F.~Wang,
Phys. Rev. D \textbf{104}, no.11, 116019 (2021)
[arXiv:2111.12307 [hep-ph]].

\bibitem{Li:2024rqb}
Y.~Li, S.~W.~Liu, E.~Wang, D.~M.~Li, L.~S.~Geng and J.~J.~Xie,
Phys. Rev. D \textbf{110}, no.7, 074010 (2024)
[arXiv:2406.01209 [hep-ph]].

\bibitem{Close:2002zu}
F.~E.~Close and N.~A.~Tornqvist,
J. Phys. G \textbf{28}, R249-R267 (2002)
[arXiv:hep-ph/0204205 [hep-ph]].

\bibitem{Debastiani:2016ayp}
V.~R.~Debastiani, W.~H.~Liang, J.~J.~Xie and E.~Oset,
Phys. Lett. B \textbf{766}, 59-64 (2017)
[arXiv:1609.09201 [hep-ph]].

\bibitem{Liang:2016hmr}
W.~H.~Liang, J.~J.~Xie and E.~Oset,
Eur. Phys. J. C \textbf{76}, no.12, 700 (2016)
[arXiv:1609.03864 [hep-ph]].

\bibitem{Wang:2021naf}
J.~Y.~Wang, M.~Y.~Duan, G.~Y.~Wang, D.~M.~Li, L.~J.~Liu and E.~Wang,
Phys. Lett. B \textbf{821}, 136617 (2021)
[arXiv:2105.04907 [hep-ph]].

\bibitem{Feng:2020jvp}
X.~C.~Feng, L.~L.~Wei, M.~Y.~Duan, E.~Wang and D.~M.~Li,
Phys. Lett. B \textbf{846}, 138185 (2023)
[arXiv:2009.08600 [hep-ph]].

\bibitem{Wang:2020pem}
Z.~Wang, Y.~Y.~Wang, E.~Wang, D.~M.~Li and J.~J.~Xie,
Eur. Phys. J. C \textbf{80}, no.9, 842 (2020)
[arXiv:2004.01438 [hep-ph]].

\bibitem{Godfrey:1985xj}
S.~Godfrey and N.~Isgur,
Phys. Rev. D \textbf{32}, 189-231 (1985)

\bibitem{Morgan:1993td}
D.~Morgan and M.~R.~Pennington,
Phys. Rev. D \textbf{48}, 1185-1204 (1993)

\bibitem{Tornqvist:1995ay}
N.~A.~Tornqvist and M.~Roos,
Phys. Rev. Lett. \textbf{76}, 1575-1578 (1996)
[arXiv:hep-ph/9511210 [hep-ph]].

\bibitem{Lee:2022jjn}
H.~J.~Lee,
New Phys. Sae Mulli \textbf{72}, no.12, 887-892 (2022)

\bibitem{Jaffe:1976ig}
R.~L.~Jaffe,
Phys. Rev. D \textbf{15}, 267 (1977).

\bibitem{Zou:1994ea}
B.~S.~Zou and D.~V.~Bugg,
Phys. Rev. D \textbf{50}, 591-594 (1994).

\bibitem{Janssen:1994wn}
G.~Janssen, B.~C.~Pearce, K.~Holinde and J.~Speth,
Phys. Rev. D \textbf{52}, 2690-2700 (1995)
[arXiv:nucl-th/9411021 [nucl-th]].

\bibitem{Oller:1997ti}
J.~A.~Oller and E.~Oset,
Nucl. Phys. A \textbf{620}, 438-456 (1997)
[erratum: Nucl. Phys. A \textbf{652}, 407-409 (1999)]
[arXiv:hep-ph/9702314 [hep-ph]].

\bibitem{Locher:1997gr}
M.~P.~Locher, V.~E.~Markushin and H.~Q.~Zheng,
Eur. Phys. J. C \textbf{4}, 317-326 (1998)
[arXiv:hep-ph/9705230 [hep-ph]].

\bibitem{Baru:2003qq}
V.~Baru, J.~Haidenbauer, C.~Hanhart, Y.~Kalashnikova and A.~E.~Kudryavtsev,
Phys. Lett. B \textbf{586}, 53-61 (2004)
[arXiv:hep-ph/0308129 [hep-ph]].

\bibitem{Albuquerque:2023bex}
R.~Albuquerque, S.~Narison and D.~Rabetiarivony,
Nucl. Phys. A \textbf{1039}, 122743 (2023)
[arXiv:2305.02421 [hep-ph]].

\bibitem{Oset:1997it}
E.~Oset and A.~Ramos,
Nucl. Phys. A \textbf{635}, 99-120 (1998)
[arXiv:nucl-th/9711022 [nucl-th]].

\bibitem{Toledo:2020zxj}
G.~Toledo, N.~Ikeno and E.~Oset,
Eur. Phys. J. C \textbf{81}, no.3, 268 (2021)
[arXiv:2008.11312 [hep-ph]].

\bibitem{Oller:2000fj}
J.~A.~Oller and U.~G.~Meissner,
Phys. Lett. B \textbf{500}, 263-272 (2001)
[arXiv:hep-ph/0011146 [hep-ph]].

\bibitem{Guo:2005wp}
F.~K.~Guo, R.~G.~Ping, P.~N.~Shen, H.~C.~Chiang and B.~S.~Zou,
Nucl. Phys. A \textbf{773}, 78-94 (2006)
[arXiv:hep-ph/0509050 [hep-ph]].

\bibitem{LHCb:2019tdw}
R.~Aaij \textit{et al.} [LHCb],
JHEP \textbf{04}, 063 (2019)
[arXiv:1902.05884 [hep-ex]].

\bibitem{BESIII:2020pxp}
M.~Ablikim \textit{et al.} [BESIII],
Phys. Rev. Lett. \textbf{124}, no.24, 241803 (2020)
[arXiv:2004.13910 [hep-ex]].

\bibitem{BaBar:2010nhz}
P.~del Amo Sanchez \textit{et al.} [BaBar],
Phys. Rev. Lett. \textbf{105}, 081803 (2010)
[arXiv:1004.5053 [hep-ex]].

\bibitem{Belle:2021dfa}
L.~K.~Li \textit{et al.} [Belle],
JHEP \textbf{09}, 075 (2021)
[arXiv:2106.04286 [hep-ex]].

\bibitem{BESIII:2019xhl}
M.~Ablikim \textit{et al.} [BESIII],
Phys. Rev. D \textbf{101}, no.5, 052009 (2020)
[arXiv:1912.12411 [hep-ex]].

\bibitem{CLEO:2008icw}
M.~Artuso \textit{et al.} [CLEO],
Phys. Rev. D \textbf{77}, 092003 (2008)
[arXiv:0802.2664 [hep-ex]].

\bibitem{BESIII:2024tpv}
M.~Ablikim \textit{et al.} [BESIII],
Phys. Rev. D \textbf{110}, no.11, L111102 (2024)
[arXiv:2404.09219 [hep-ex]].

\bibitem{Wang:2021kka}
Z.~Y.~Wang, H.~A.~Ahmed and C.~W.~Xiao,
Phys. Rev. D \textbf{105}, no.1, 016030 (2022)
[arXiv:2110.05359 [hep-ph]].

\bibitem{Liang:2014ama}
W.~H.~Liang, J.~J.~Xie and E.~Oset,
Phys. Rev. D \textbf{92}, no.3, 034008 (2015)
[arXiv:1501.00088 [hep-ph]].

\bibitem{Xie:2014gla}
J.~J.~Xie and E.~Oset,
Phys. Rev. D \textbf{90}, no.9, 094006 (2014)
[arXiv:1409.1341 [hep-ph]].

\bibitem{Ling:2021qzl}
X.~Z.~Ling, M.~Z.~Liu, J.~X.~Lu, L.~S.~Geng and J.~J.~Xie,
Phys. Rev. D \textbf{103}, no.11, 116016 (2021)
[arXiv:2102.05349 [hep-ph]].

\bibitem{Ding:2024lqk}
Y.~Ding, E.~Wang, D.~M.~Li, L.~S.~Geng and J.~J.~Xie,
Phys. Rev. D \textbf{110}, no.1, 014032 (2024)
[arXiv:2401.17322 [hep-ph]].

\bibitem{Ding:2023eps}
Y.~Ding, X.~H.~Zhang, M.~Y.~Dai, E.~Wang, D.~M.~Li, L.~S.~Geng and J.~J.~Xie,
Phys. Rev. D \textbf{108}, no.11, 114004 (2023)
[arXiv:2306.15964 [hep-ph]].

\bibitem{Dai:2018rra}
L.~R.~Dai, Q.~X.~Yu and E.~Oset,
Phys. Rev. D \textbf{99}, no.1, 016021 (2019)
[arXiv:1809.11007 [hep-ph]].

\bibitem{BESIII:2025wmd}
M.~Ablikim \textit{et al.} [BESIII],
[arXiv:2512.23389 [hep-ex]].

\bibitem{BESIII:2025sea}
M.~Ablikim \textit{et al.} [BESIII],
[arXiv:2510.25111 [hep-ex]].

\bibitem{CLEO:2004umu}
P.~Rubin \textit{et al.} [CLEO],
Phys. Rev. Lett. \textbf{93}, 111801 (2004)
[arXiv:hep-ex/0405011 [hep-ex]].

\bibitem{CLEO:2011cnt}
N.~Lowrey \textit{et al.} [CLEO],
Phys. Rev. D \textbf{84}, 092005 (2011)
[arXiv:1106.3103 [hep-ex]].

\bibitem{Xie:2014tma}
J.~J.~Xie, L.~R.~Dai and E.~Oset,
Phys. Lett. B \textbf{742}, 363-369 (2015)
[arXiv:1409.0401 [hep-ph]].

\bibitem{Zhang:2024myn}
X.~H.~Zhang, H.~Zhang, B.~C.~Ke, L.~J.~Liu, D.~M.~Li and E.~Wang,
Phys. Rev. D \textbf{110}, no.11, 114050 (2024)
[arXiv:2409.09966 [hep-ph]].

\bibitem{Ikeno:2024fjr}
N.~Ikeno, J.~M.~Dias, W.~H.~Liang and E.~Oset,
Eur. Phys. J. C \textbf{84}, no.5, 469 (2024)
[arXiv:2402.04073 [hep-ph]].

\bibitem{BESIII:2023htx}
M.~Ablikim \textit{et al.} [BESIII],
Phys. Rev. Lett. \textbf{132}, no.13, 131903 (2024)
[arXiv:2309.05760 [hep-ex]].

\bibitem{Belle:2020fbd}
Y.~Q.~Chen \textit{et al.} [Belle],
Phys. Rev. D \textbf{102}, no.1, 012002 (2020)
[arXiv:2003.07759 [hep-ex]].

\bibitem{MartinezTorres:2009uk}
A.~Martinez Torres, L.~S.~Geng, L.~R.~Dai, B.~Sun, Xi, E.~Oset and B.~S.~Zou,
Phys. Lett. B \textbf{680}, 310-315 (2009)
[arXiv:0906.2963 [nucl-th]].

\bibitem{Bramon:1992kr}
A.~Bramon, A.~Grau and G.~Pancheri,
Phys. Lett. B \textbf{283}, 416-420 (1992)

\bibitem{Roca:2003uk}
L.~Roca, J.~E.~Palomar and E.~Oset,
Phys. Rev. D \textbf{70}, 094006 (2004)
[arXiv:hep-ph/0306188 [hep-ph]].

\bibitem{Gamermann:2009ouq}
D.~Gamermann, E.~Oset and B.~S.~Zou,
Eur. Phys. J. A \textbf{41}, 85-91 (2009)
[arXiv:0805.0499 [hep-ph]].

\bibitem{Molina:2019udw}
R.~Molina, J.~J.~Xie, W.~H.~Liang, L.~S.~Geng and E.~Oset,
Phys. Lett. B \textbf{803}, 135279 (2020)
[arXiv:1908.11557 [hep-ph]].

\bibitem{Liang:2014tia}
W.~H.~Liang and E.~Oset,
Phys. Lett. B \textbf{737}, 70-74 (2014)
[arXiv:1406.7228 [hep-ph]].

\bibitem{Wang:2021ews}
Z.~Y.~Wang, J.~Y.~Yi, Z.~F.~Sun and C.~W.~Xiao,
Phys. Rev. D \textbf{105}, no.1, 016025 (2022)
[arXiv:2109.00153 [hep-ph]].

\bibitem{Ahmed:2020qkv}
H.~A.~Ahmed, Z.~Y.~Wang, Z.~F.~Sun and C.~W.~Xiao,
Eur. Phys. J. C \textbf{81}, no.8, 695 (2021)
[arXiv:2011.08758 [hep-ph]].

\bibitem{Liang:2023ekj}
W.~Liang, J.~Y.~Yi, C.~W.~Xiao and S.~Q.~Zhou,
Phys. Rev. D \textbf{108}, no.11, 116001 (2023)

\bibitem{Dai:2021owu}
L.~R.~Dai, E.~Oset and L.~S.~Geng,
Eur. Phys. J. C \textbf{82}, no.3, 225 (2022)
[arXiv:2111.10230 [hep-ph]].

\bibitem{Oller:1998zr}
J.~A.~Oller and E.~Oset,
Phys. Rev. D \textbf{60}, 074023 (1999)
[arXiv:hep-ph/9809337 [hep-ph]].

\bibitem{Gamermann:2006nm}
D.~Gamermann, E.~Oset, D.~Strottman and M.~J.~Vicente Vacas,
Phys. Rev. D \textbf{76}, 074016 (2007)
[arXiv:hep-ph/0612179 [hep-ph]].

\bibitem{Lin:2021isc}
J.~X.~Lin, J.~T.~Li, S.~J.~Jiang, W.~H.~Liang and E.~Oset,
Eur. Phys. J. C \textbf{81}, no.11, 1017 (2021)
[arXiv:2109.08405 [hep-ph]].

\bibitem{Roca:2020lyi}
L.~Roca and E.~Oset,
Phys. Rev. D \textbf{103}, no.3, 034020 (2021)
[arXiv:2011.05185 [hep-ph]].














\end{thebibliography}
\end{document}